\documentclass[journal=jctc,manuscript=article]{achemso}
\usepackage[utf8]{inputenc}
\usepackage[T1]{fontenc}
\usepackage[version=3]{mhchem}
\usepackage{soul}
\usepackage{bm}
\usepackage{threeparttable}
\usepackage{color}


\title{A $\Delta$-Machine Learning Approach for Force Fields, Illustrated by a CCSD(T) 4-body Correction to the MB-pol Water Potential}

\author{Chen Qu}
\affiliation{Independent Researcher, Toronto, Canada}
\email{szquchen@gmail.com}
\author{Qi Yu}
\affiliation{Department of Chemistry Yale University, New Haven, Connecticut 06520, U.S.A.}
\author{Riccardo Conte}
\affiliation{Dipartimento di Chimica, Universit\`{a} degli Studi di Milano, via Golgi 19, 20133 Milano, Italy}
\author{Paul L. Houston}
\affiliation{Department of Chemistry and Chemical Biology, Cornell University, Ithaca, New York
14853, U.S.A. and Department of Chemistry and Biochemistry, Georgia Institute of
Technology, Atlanta, Georgia 30332, U.S.A}
\author{Apurba Nandi}
\affiliation{Department of Chemistry and Cherry L. Emerson Center for Scientific Computation, Emory University, Atlanta, Georgia 30322, U.S.A.}
\author{Joel M. Bowman}
\email{jmbowma@emory.edu}
\affiliation{Department of Chemistry and Cherry L. Emerson Center for Scientific Computation, Emory University, Atlanta, Georgia 30322, U.S.A.}

\begin{document}

%\keywords{Keyword1, Keyword2, Keyword3}
\newpage
\begin{abstract}
%The TTM4-F potential is a recent version of the family of polarizable TTMn-F %potentials for water.  These potentials fail to describe the strong interactions %in the short range where monomer electron densities overlap.  
$\Delta$-Machine Learning ($\Delta$-ML) has been shown to effectively and efficiently bring a low-level ML potential energy surface to CCSD(T) quality.  Here we propose extending this approach to general force fields, which implicitly or explicitly contain many-body effects.  After describing this general approach, we illustrate it for the MB-pol water potential which contains CCSD(T) 2-body and 3-body interactions but relies on the TTM4-F 4-body and higher body interactions.  The 4-body MB-pol (TTM4-F) interaction fails at very short range and for the water hexamer errors up to 0.84 kcal/mol are seen for some isomers, owing mainly to 4-body errors. We apply $\Delta$-ML for the 4-body interaction, using a recent dataset of CCSD(T) 4-body energies that we used to develop a new water potential, q-AQUA.  This 4-body correction is shown to improve the accuracy of the MB-pol potential for the relative energies of 8 isomers of the water hexamer as well as the harmonic frequencies. The new potential is robust in the very short range and so should be reliable for simulations at high pressure and/or high temperature.  
\end{abstract}

\flushbottom
\maketitle

\thispagestyle{empty}
\newpage
\section*{Introduction}

Machine learning (ML) correction methods aim to elevate the level of accuracy of ML properties, for example potential energy surfaces (PESs). There are two approaches currently being investigated to accomplish this goal.  One is transfer learning, which has been developed extensively in the context of artificial neural networks,\cite{TL_ieee} and much of the work in that field has been brought into chemistry, especially in the development of PESs.\cite{roit19, meuwly20, TL_2020} For example, Meuwly and co-workers applied transfer learning using thousands of local CCSD(T) energies to improve their MP2-based neural network PESs for malonaldehyde, acetoacetaldehyde and acetylacetone.\cite{meuwly20}
The basic idea of transfer learning is that a fit obtained from one source of data (perhaps a large one) can be fine-tuned for a related problem by using limited data. Therefore, in the present context of PES fitting, an ML-PES trained with low-level electronic energies/gradients can be reused as the starting point of the model for an ML-PES with the accuracy of a high-level electronic structure theory. As noted, this is typically done with artificial neural networks, where weights and biases trained on lower-level data hopefully require minor changes in response to additional training using high-level data. 

The other approach is $\Delta$-machine learning.  In this approach a correction is made to a property obtained using an efficient, low-level \textit{ab initio} theory.\cite{Lilienfeld15, Stohr2020, Tuckerman_ML_2020, Csanyi_DeltaML} 
We recently proposed and tested a $\Delta$-ML approach, that uses a small number of CCSD(T) energies, to correct a low-level PES based on DFT electronic energies and gradients.\cite{NandiDeltaML2021, QuDelteMLAcAc2021} The equation for this approach is simply
\begin{equation} 
    V_{LL{\rightarrow}CC}=V_{LL}+\Delta{V_{CC-LL}},
\end{equation}
where $V_{LL{\rightarrow}CC}$ is the corrected PES,  $V_{LL}$ is a PES fit to low-level electronic data (such as DFT energies and gradients), and $\Delta{V_{CC-LL}}$ is the correction based on high-level coupled cluster energies. It is noted that the difference between CCSD(T) and DFT energies, $\Delta{V_{CC-LL}}$, is usually not as strongly varying as $V_{LL}$ with respect to the nuclear configurations and therefore just a small number of high-level electronic energies are adequate to fit the correction PES. The method was validated for PESs of small molecules, \ce{CH4} and \ce{H3O+}, 12-atom $N$-methyl acetamide, and 15-atom acetylacetone. In all cases, the coupled cluster energies were obtained over the same large span of configurations used to get the lower-level PES.

Here we propose to extend this $\Delta$-ML approach from molecular PESs to general, non-reactive force fields that are explicitly or implicitly many-body.  There are many examples of such force fields that determine the total energy of $N$ monomers.  For example, consider force fields for water.  (For a recent review see ref \citenum{mbreview}.)  For simplicity, we denote these by ``MB-FF''.  Suppose our goal is to bring this force field to the ``gold-standard'' CCSD(T) level of theory.  Clearly this cannot be done by simply applying the above equation for an arbitrary number of monomers owing to the prohibitively exponential computational cost of CCSD(T) calculations with respect to the number of monomers.  Instead we propose a $\Delta$-ML force-field for $N$ monomers given by the sum of many-body corrections, namely 
\begin{equation}
  V_\text{$\Delta$-ML} = V_\text{MB-FF} + \sum_{i>j}^N\Delta{V_{2-b}(i,j)}+\sum_{i>j>k}^N\Delta{V_{3-b}}(i,j,k)+\sum_{i>j>k>l}^N\Delta{V_{4-b}}(i,j,k,l) + \cdots,
\end{equation}
where $\Delta V_{n-b}$ are the many-body corrections to the MB-FF many-body terms., given by the difference between CCSD(T) and MB-FF $n$-body ($n$-b) interaction energies. To be clear, recall that the $n$-b interaction energy is obtained from a cluster of $n$ monomers. For example, the 4-b interaction is obtained by calculating the total energy of the tetramer (four monomers) and subtracting all the 1-, 2-, and 3-b interactions from the total energy. Note for simplicity, we assume that an accurate 1-b term, e.g., the single water molecule, is given in the MB-FF.

We have truncated explicit correction terms at the 4-b level with force-fields for water in mind. This is because it has been established by high-level calculations that 4-b, while small, are needed to obtain nearly 100 percent of the electronic dissociation energies of water clusters up to the 21-mer.\cite{MBE20} This has also been shown previously for some isomers of the water hexamer\cite{KShexamer, mbpoltests} and we show this again explicitly here. And crucially, we have the CCSD(T) electronic energies needed for the correction up to 4-b, and these datasets are recently used to develop a pure many-body water potential q-AQUA.\cite{4bjpcl21, q_AQUA}
\begin{equation}
V_\text{q-AQUA} = \sum_{i=1}^NV_{1-b}(i) + \sum_{i>j}^NV_{2-b}(i,j) +  \sum_{i>j>k}^NV_{3-b}(i,j,k) +\\
 \sum_{i>j>k>l}^NV_{4-b}(i,j,k,l),
\end{equation}
where the meaning of each term is clear. In q-AQUA the 2-, 3-, and 4-b interactions are permutationally invariant polynomial (PIP) \cite{braams09, Xie10} fits to thousands of CCSD(T) energies (details are in ref. \citenum{q_AQUA}.) These CCSD(T) electronic energies for the 2-b, 3-b and 4-b are ready to be used to obtain a $\Delta V_{n-b}$ correction to any water MB-FF, and we return to this below.

%These were fit using  permutationally invariant polynomials (PIPs)\cite{braams09, msachen, purified15c}. As for the 2-b and 3-b interactions, we recently reported those at the CCSD(T) level, as well as an extended 4-b interaction energies in our new water potential q-AQUA.\cite{q_AQUA}  That potential is given by a pure many-body expansion.

%but with a difference in the specifics.  Namely, the correction is to a polarizable force field, specifically the 4-b component of TTM4-F. We have also retroactively characterized the MB-pol strategy in terms of the general $\Delta$-ML approach.  
%There are now several potential energy surfaces (PESs) for flexible water based on fits to high-level \textit{ab initio} data for 2 and 3-body interactions and a spectroscopically accurate PES for the water monomer.\cite{WHBB, Babin2012, ccflex23} The WHBB\cite{WHBB} and MB-pol\cite{Babin2012} PESs are augmented by the so-called TTMn-F family of potentials. The WHBB PES\cite{WHBB} uses TTM3-F\cite{TTM3F} and the MB-pol PES \cite{Babin2012} uses TTM4-F.\cite{TTM4} These are sophisticated potentials that describe the long-range interaction of an arbitrary number of water monomers.  They are not accurate in the short-range and so are replaced there by fits to tens of thousand of high-level electronic energies using permutationally invariant polynomials,\cite{Braams2009} both in WHBB and MB-pol. There are significant differences in the implementation of this general approach in these two PESs. 

In this work, the focus is on the 4-b correction to the MB-pol force field.\cite{mbpol2b, mbpol3b}  In MB-pol the 2-b and 3-b are already at CCSD(T) level, but the 4-b interaction is essentially given by the TTM4-F potential,\cite{TTM4} which is a sophisticated MB-FF for water. Errors between 0.1 and 0.84 kcal/mole for the these 4-b interactions for the hexamer isomers against direct CCSD(T) calculations, were reported in 2015.\cite{medders15} These are fairly large fractions of the 4-b energy itself. Stimulated by recent assessments of the importance of the 4-b interaction and inaccuracy of the MB-pol 4-b, we report a correction 4-b PES, denoted $\Delta V_{4-b}$, that is aimed directly at extending the MB-pol potential to the CCSD(T) 4-b level. The correction is a PIP fit to the energy difference between the CCSD(T) 4-b interactions and TTM4-F 4-b interactions.

In the next section we present the details of the 4-b correction PES, followed by several tests that indicate the effectiveness of this correction PES.

\section*{$\Delta V_{4-b}$ Fitting Details}

The data set for the $\Delta V_{4-b}$ fit is simply the difference between the 4-b CCSD(T)-F12/haTZ (aug-cc-pVTZ basis for O atoms and cc-pVTZ for H atoms) and MB-pol/TTM4-F energies. The total number of configurations used for this fit is 3695. The fit uses the PIP approach, in which the PIPs are generated using MSA software.\cite{Xie10, msachen} The PIPs are usually polynomials in the Morse variables of the internuclear distances, $\exp(-r_{ij}/\lambda)$, where $r_{ij}$ is the distance between atoms $i$ and $j$ and $\lambda$ is a range parameter, taken for this calculation to be 1.5 bohr.  We used 22221111 permutational symmetry at a maximum polynomial order of 3. Two additional issues concerning the basis set are important to note.

First, it is desirable not to include polynomials that do not have the correct limiting behavior as one or more monomers are removed from the others to a large distance.  In the 4-body case, we need to consider the removal of each monomer from the other three and the removal of each possible dimer from the other one.  In all of these cases, the 4-body interaction energy must vanish.  The process of identifying PIPs that do not have the correct limiting behavior is what we call purification.\cite{purified13, purified14} To identify the PIPs with incorrect limiting behavior, the relevant distances are augmented by 100 \AA, and we accept the polynomial as having the correct behavior if its Morse value is below $10^{-6}$.  We cannot, however, immediately eliminate these polynomials because there may be other polynomials that, for example, are composed of products between one with a correct limit and one with an incorrect limit.  At first, we simply rename the ones with an incorrect limit.  After all the polynomials have been evaluated, we examine the definitions of all those with the correct limits and determine which of the monomials and which of the renamed polynomials with incorrect limits are required to calculate them.  Finally, we remove those polynomials that are not required and renumber those that remain, keeping the order of calculation to ensure that no partial calculation that contributes to any polynomial needs to be performed twice.  We then have a set of polynomials that all have the correct limiting behavior and that can be calculated efficiently.\cite{conte20}

The second issue that we need to consider is how to maintain permutational symmetry, not only in each monomer, but when monomers as a whole are interchanged with one another. This latter exchange is not taken into account by the MSA software, so the polynomials that we create by the previously described purification will not, in general, have permutational symmetry with respect to exchange of identical monomers.  A common method for dealing with this issue is to augment the dataset by adding all relevant permutations of the Cartesian coordinates and assigning them the same energy, thus requiring a set of $n!$ geometries for each energy, where $n$ is the number of monomers (4, in this case). A better method is to identify groups of polynomials that have permutational symmetry with respect to monomer exchange and then form ``superpolynomials'' that are the sum of the polynomial members of each group. We identify the permutationally invariant groups of polynomials by taking a single set of $n!$ permutationally related geometries and calculating the value of each polynomial.  While the the values of individual polynomials vary from permutation to permutation, the groups of polynomials, taken together for each permutation, will have the same group of values. For each permutation, one can form pairs of the polynomial identities and their values, and then sort the pairs by their values.  Looking at all pairs that have the same value component in all permutations gives the identities of the polynomials, some of which may be repeated, that make up a permutationally invariant group. In general, there will be as many groups as there were original polynomials. These groups, each with $n!$ (not necessarily unique) polynomial contributions, are then summed to form ``superpolynomials'' having permutational symmetry with respect to exchange of identical molecules. Having formed these superpolynomials, there is no need for augmentation of the dataset with permutationally related geometries. 

We used basis sets of different sizes, with 200, 500, and 1000 ``superpolynomials''. More details of the bases are given in Table \ref{tab:basis}. As seen, although fitting with more polynomials can reduce the fitting error, the computational cost is roughly proportional to the sum of the number of monomials and polynomials. The results presented in this paper are based on the basis with 200 ``super-polynomials'', as this achieves reasonably good accuracy with smaller cost.

The final energy is written as
\begin{equation}
    E = E_\text{MB-pol} + \sum_{i>j>k>l} S_{ijkl} \Delta V_{4-b}(i,j,k,l),
\end{equation}
where $S_{ijkl}$ is a switching function whose value is 1 at short range and 0 at the long range. Specifically,
\begin{align}
    S = 10 \left(\frac{r_\text{max} - r_i}{r_f - r_i}\right)^3 - 15 \left(\frac{r_\text{max} - r_i}{r_f - r_i}\right)^4 + 6 \left(\frac{r_\text{max} - r_i}{r_f - r_i}\right)^5~~~~~(r_i < r_\text{max} < r_f),
\end{align}
where $r_\text{max}$ is the maximum OO distance in a water tetramer, and $S$ is 1 when $r_\text{max}$ is smaller than $r_i$ and is 0 when $r_\text{max}$ is greater than $r_f$. In this work we used $r_i = 5.5$ \AA~ and $r_f = 7.0$ \AA~ unless otherwise specified.

\begin{table}[htbp!]
    \centering
	\begin{tabular*}{0.5\columnwidth}{@{\extracolsep{\fill}}lccc}
	\hline
	\hline\noalign{\smallskip}
      & PIP$_{200}$ & PIP$_{500}$ & PIP$_{1000}$  \\
	\noalign{\smallskip}\hline\noalign{\smallskip}
	 Number of m & 1438 &  3442 &  8610 \\
	 Number of q & 5490 & 12898 & 25084 \\
     Number of p &  200 &   500 &  1000 \\
    Fitting RMSE &  6.7 &   4.0 &   2.5 \\
       Timing    &  1.1 &   2.7 &   6.0 \\
	\noalign{\smallskip}\hline
	\hline
	\end{tabular*}
    \caption{Number of monomials (m), polynomials (q), and ``super-polynomials'' (p) in the three fitting bases used for $\Delta V_{4-b}$, and corresponding fitting root-mean-square error (RMSE, in cm$^{-1}$) and computational time (in seconds) for 100,000 energy evaluations. The computational time for all gradients is about 3 times that for the energy.\cite{Houston2022}}
    \label{tab:basis}
\end{table}

\section*{Results and Discussions}

\begin{figure}
    \centering
    \includegraphics[width=\columnwidth]{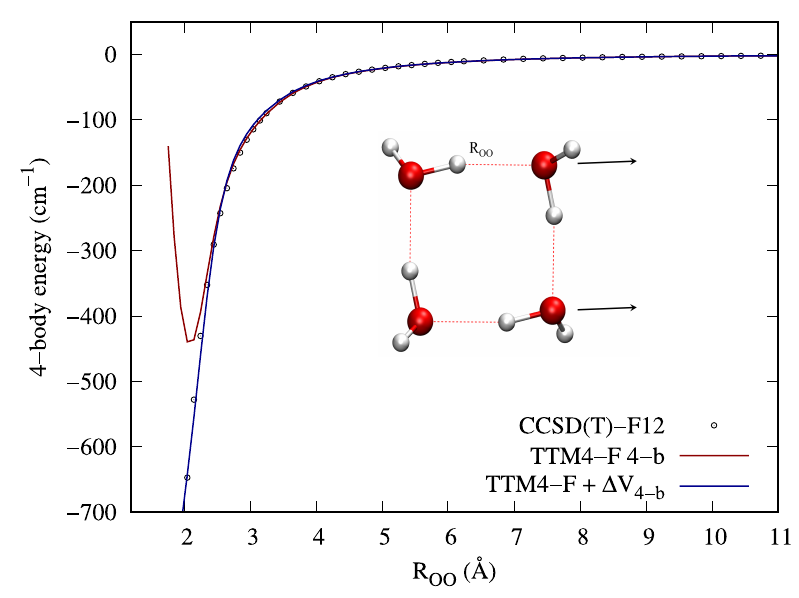}
    \caption{4-b energies from indicated sources as a function of the oxygen-oxygen distance between pairs of water dimers in the tetramer. The arrows indicate the dimer pair that separates from the rigid tetramer. The equilibrium value of this distance is 2.7 {\AA}.}
    \label{fig:22cut}
\end{figure}

\begin{figure}[htbp!]
    \centering
    \includegraphics[width=\columnwidth]{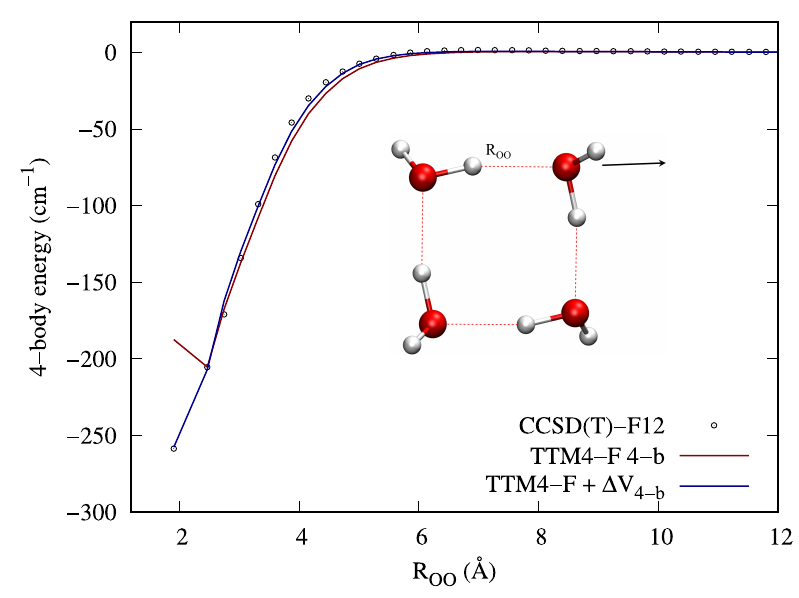}
    \caption{4-b energies from indicated sources with a single monomer separating from the tetramer. R$_\text{OO}$ is the distance between the O atoms on the two monomers on the axis inferred from the arrow.}
    \label{fig:13cut}
\end{figure}

First we examine two 1-d cuts where we compare the 4-b CCSD(T)-F12/haTZ energies to those from MB-pol/TTM4-F and to those from MB-pol+$\Delta V_{4-b}$. Figures \ref{fig:22cut} and \ref{fig:13cut} show cuts of the potential for separating two dimers from one another and for separating a monomer from the remaining trimer, respectively.  The major improvements of the 4-b correction are in the short-range. Whereas MB-pol/TTM4-F is quite accurate in the long range, it is not designed with the proper Pauli exchange and repulsion in the short range. Despite the fact that TTM4-F fails badly in the short range, the $\Delta_{ 4-b}$ potential does provide a reasonable correction.  It should be noted that for both cuts shown, the equilibrium R$_{\text{OO}}$ distance is 2.7 \AA, so that large corrections are in the steeply repulsive part of the potential. However, there are also large corrections for highly distorted tetramers geometries not visited by this cut, discussed below.

\begin{figure}
    \centering
    \includegraphics[width=\columnwidth]{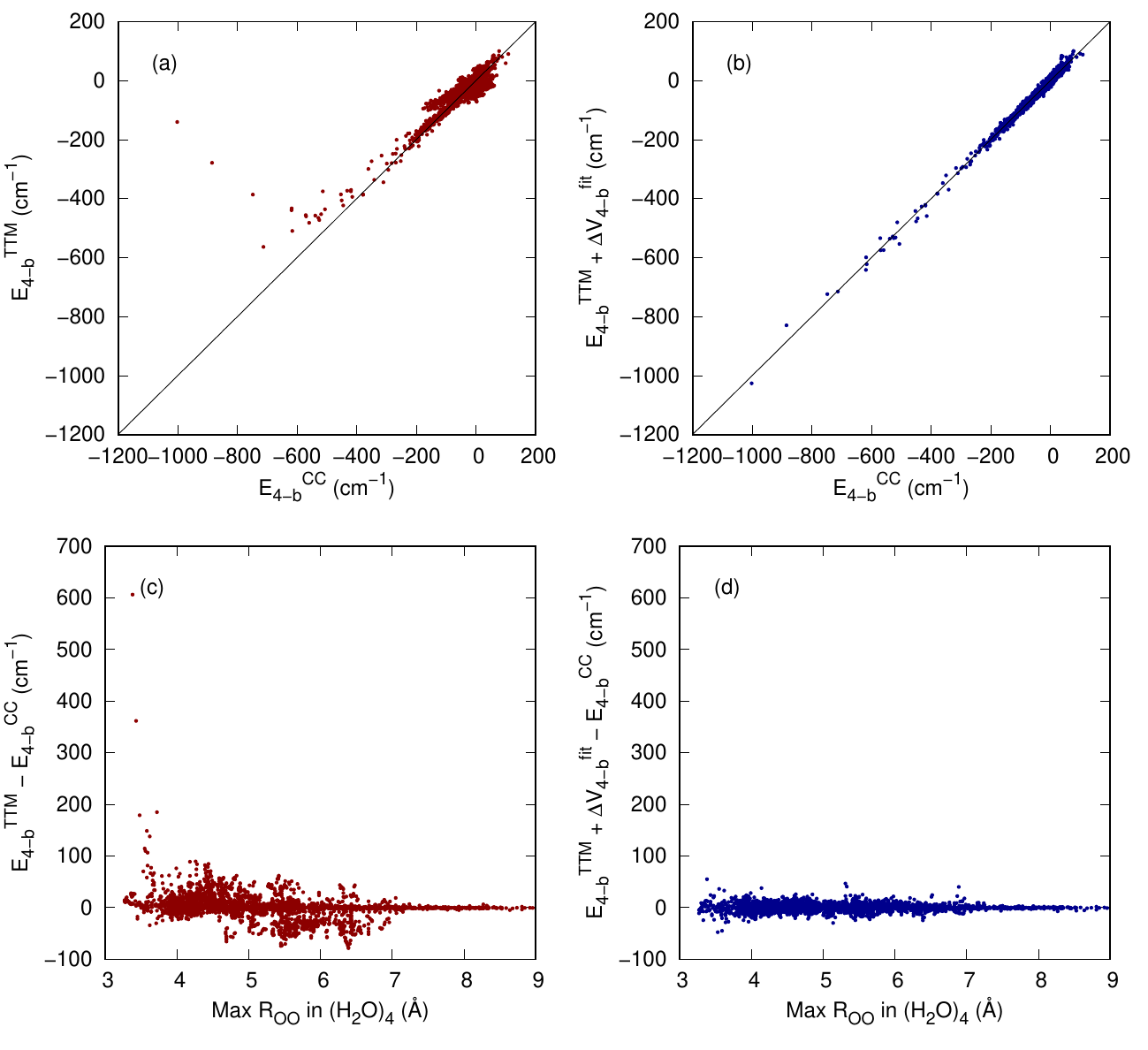}
    \caption{(a) Correlation plot between TTM4-F 4-b and CCSD(T)-F12 4-b energies; (b) correlation plot between TTM4-F+$\Delta V_{4-b}$ and CCSD(T)-F12 4-b energies; (c) error of TTM4-F 4-b as a function of max R$_\text{OO}$ in the tetramer; (d) error of TTM4-F+$\Delta V_{4-b}$ as a function of max R$_\text{OO}$ in the tetramer}
    \label{fig:error}
\end{figure}

Figure \ref{fig:error} shows in the top panels the correlation plots between the TTM4-F 4-b and CCSD(T)-F12 4-b energies (panel a), and between $E_\text{4-b}^\text{TTM4-F}+\Delta V_\text{4-b}$ and CCSD(T)-F12 4-b energies (panel b).  The bottom panels plot, as a function of the maximum R$_\text{OO}$ distance of the tetramer, the difference between TTM4-F and CCSD(T) energies (panel c) and the difference between the corrected TTM4-F and CCSD(T) energies (panel d).  It is visually clear that the correction provides both a better correlation and a smaller error with respect to the CCSD(T)-F12 4-b energies. Note that in addition to large errors in the short range, the TTM4-F 4-b energies also have significant errors (>50 cm$^{-1}$) even when the R$_\text{OO}$ reaches 6.5 \AA, as panel c shows.

\begin{figure}[htbp!]
    \centering
    \includegraphics[width=\columnwidth]{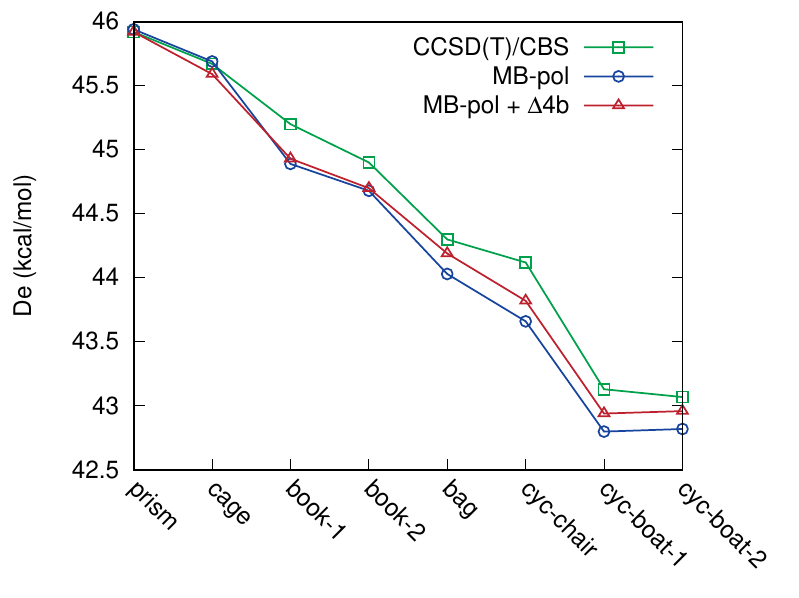}
    \caption{The binding energies of the eight isomers of the water hexamer from indicated sources.}
    \label{fig:hexamer}
\end{figure}

Then consider the binding energies of the eight isomers of the water hexamer as shown in Fig. \ref{fig:hexamer}. The benchmark CCSD(T)/CBS values are taken from ref. \citenum{bates09}. As seen in the figure, with the 4-b correction, the binding energies are in better agreement with the benchmark values, especially for bag and cyclic isomers.

Table \ref{tab:freq} shows for four of the hexamer isomers the mean absolute error (MAE) in the harmonic frequencies for both uncorrected and corrected MB-pol PES. For prism and cage, the 4-b corrected version is essentially as accurate as the original MB-pol, while for book and cyclic ring, the 4-b correction clearly improves the frequencies.  

\begin{table}[htbp!]
    \centering
    \small

	\begin{tabular*}{\columnwidth}{@{\extracolsep{\fill}}cccccccccccc}
	\hline
	\hline\noalign{\smallskip}
	  & \multicolumn{2}{c}{Prism} && \multicolumn{2}{c}{Cage} && 
	    \multicolumn{2}{c}{Book-1} && \multicolumn{2}{c}{Cyclic Ring}\\
    \noalign{\smallskip} \cline{2-3} \cline{5-6} \cline{8-9} \cline{11-12} \noalign{\smallskip}
     & MB-pol & +$\Delta V_{4-b}$ && MB-pol & +$\Delta V_{4-b}$ && MB-pol & +$\Delta V_{4-b}$ && MB-pol & +$\Delta V_{4-b}$ \\
	\noalign{\smallskip}\hline\noalign{\smallskip}
    MAE & 7.8 & 8.4 && 8.9 & 9.4 && 12.6 & 10.6 && 16.5 & 11.7 \\
	\noalign{\smallskip}\hline
	\hline
   
	\end{tabular*}
	
    \caption{Mean absolute errors (MAE) in harmonic frequencies (in cm$^{-1}$) for indicated PESs. The benchmarks are from ref. \citenum{Howard2015}.}
    \label{tab:freq}
\end{table}

Next, we comment briefly on the timing requirements for the TTM4-F+$\Delta V_{4-b}$ potential.  As shown in the last line of Table \ref{tab:basis}, the additional time for calculating the correction for the results presented in this paper (PIP$_\text{200}$) is only about 1 second for 100,000 energy evaluations and approximately 3 seconds if the energy and associated gradients are evaluated at the same time. Table \ref{tab:timing} shows the overall timing to evaluate the energy of 64, 128, and 256 monomers, with different cutoffs of the 4-b correction. In this table, $t_\text{MBX}$ is the time to evaluate all the terms in the MB-pol, including the TTM4-F and MB-pol 2-b and 3-b, with the latest MBX software,\cite{MBX} while $t_{\Delta V_{4-b}}$ is the time to evaluate our 4-b correction term, which is the extra cost when the $\Delta V_{4-b}$ is added to MB-pol. All timings are evaluated using a single Intel i7-8750H core.
It can be seen that the extra cost of the 4-b correction is in general of the same order of magnitude as the cost of MB-pol, and the cutoff distance can be tuned to achieve a balance between the cost and accuracy.

\begin{table}[htbp!]
    \centering

	\begin{tabular*}{\columnwidth}{@{\extracolsep{\fill}}cccccccccc}
	\hline
	\hline\noalign{\smallskip}
	  &  & \multicolumn{2}{c}{Cutoff = 6.0 \AA} && \multicolumn{2}{c}{Cutoff = 6.5 \AA} && \multicolumn{2}{c}{Cutoff = 7.0 \AA} \\
    \noalign{\smallskip} \cline{3-4} \cline{6-7} \cline{9-10} \noalign{\smallskip}
     $N_\text{mono}$ & $t_\text{MBX}$ & $N_\text{tetra}$ & $t_{\Delta V_{4-b}}$ && $N_\text{tetra}$ & $t_{\Delta V_{4-b}}$ && $N_\text{tetra}$ & $t_{\Delta V_{4-b}}$ \\
	\noalign{\smallskip}\hline\noalign{\smallskip}
    64 & 0.057 & 1379 & 0.011 && 2728 & 0.019 && 5639 & 0.035 \\
    128 & 0.21 & 5823 & 0.085 && 12224 & 0.13 && 24460 & 0.19 \\
    256 & 0.68 & 13023 & 1.03 && 28786 & 1.15 && 58804 & 1.32 \\
	\noalign{\smallskip}\hline
	\hline
   
	\end{tabular*}
	
    \caption{Time (in seconds) needed to evaluate the energy of $N_\text{mono}$ monomers. Here $t_\text{MBX}$ is the time to obtain all terms in MB-pol using the latest version (MBX), and $t_{\Delta V_{4-b}}$ is the time to evaluate our 4-b correction. $N_\text{tetra}$ is the number of tetramers needed to be evaluated.}
    \label{tab:timing}
\end{table}

We have made comparisons between the TTM4-F 4-b potential, the TTM4-F 4-b with $\Delta V_{4-b}$ correction, and the CCSD(T) results. A remaining question that should be addressed is how the correction potential performs in comparison with the previously reported full 4-b potential\cite{4bjpcl21} as recently improved.\cite{q_AQUA}  Of course, many larger studies may already be based on the very successful MB-pol potential.  For these, improvement by $\Delta V_{4-b}$ might be the easiest upgrade, with some extra computational cost, depending on the choice of the 4-b cutoff. But for those who are interested only in the 4-b potential, we suggest previously reported 4-b PES since it is much faster than TTM4-F 4-b $+\Delta V_{4-b}$, and is slightly more accurate as well. 

Finally, we note that many potentials or components of potentials can be corrected by this method, which has already been shown to be accurate and efficient for \ce{CH4}, \ce{H3O+}, NMA,\cite{NandiDeltaML2021} and for AcAc,\cite{QuDelteMLAcAc2021} with more applications in process.  In the current study, we have shown substantial improvement of the 4-b potential for TTM4-F, but one might have reasonable hope that this $\Delta$-ML  method is a general approach that could provide substantial improvements to other potentials at relatively small cost. In this case, the corrections would begin with the 2-b ones and could go up to 4-b. Our recent CCSD(T) 2-b, 3-b, and 4-b datasets are available,\cite{datasets} and so the corrections, $\Delta V_{n-b}$, simply require a water force field.  Some examples are the well-known potentials AMOEBA,\cite{amoeba13} and TTM2.1\cite{TTM2F} or TTM3.\cite{TTM3F} These are polarizable potentials, however, with significant differences.  Another force field that might be interesting to ``correct'' is MB-UCB \cite{MBUCB}. Since this potential relies heavily on DFT calculation, using the $\omega$B97X-V/def2-QZVPPD functional, it appears that the correction to MB-UCB would analogous to the correction of DFT to CCSD(T) PESs that we considered originally our Delta-ML method. 

\section*{Summary and Conclusions}

The 4-b interaction in the MB-pol potential has been corrected using the proposed $\Delta$-ML approach. The 4-b interaction itself and the correction are ``small'' compared to say the 2-b interaction.  However, as noted above the 4-b correction has been shown to be the ``ultimate'' interaction by Xantheas and co-workers for large water clusters.  Further it extends the successful MB-pol potential to this level of interaction.  It is worth noting that the PIP fit to the difference 4-b energies is challenging because it is a 12-atom PES. While this number of atoms is not at the frontier of ML-PESs currently, it was not feasible years ago when PIP 2-b and 3-b potentials were reported for water in the WHBB \cite{WHBB} and the MB-pol\cite{mbpol2b, mbpol3b} potentials.

That the correction potential is a significant improvement over the TTM4-F potential can be seen a) by comparing cuts of the potentials for TTM4-F 4-b, and TTM4-F 4-b + $\Delta V_{4-b}$ along with the CCSD(T)-F12 values; b) by comparing the correlation between these potentials with the CCSD(T) values and the errors as a function of R$_{\text{OO}}$; c) by comparing results for the binding energies of the water hexamer isomers; and d) comparing the MAEs in harmonic frequencies for four of the isomers.

The methods described above offer three important ideas. First, we have described a method for ensuring that the potentials go to zero when appropriate distances get large. Second, we have also described a method for maintaining permutational symmetry not only in each monomer but when monomers as a whole are interchanged with one another. Finally, the $\Delta V$ method itself allows large improvements for small amount of effort, and this approach appears to be general and could be applied for other water force fields and similar types of force fields for other liquids and materials.

\section*{Acknowledgment}
JMB thanks the ARO, DURIP grant (W911NF-14-1-0471), for funding a computer cluster where most of the calculations were performed and current financial support from NASA (80NSSC20K0360). We are thankful for correspondence with Markus Meuwly and Silvan Käser.

%\section*{Methods}

%\subsection*{Diffusion Monte Carlo Calculations}

%\subsection*{One-dimensional Tunneling Calculation}

%\section*{Summary and Conclusions}
%We described a full-dimensional potential energy surface (PES) for acetylacetone using a recently reported $\Delta$-machine learning, permutationally invariant polynomial approach. The approach is applied to a recent MP2-based PES, which although realistic, has a barrier height for H-atom transfer that is low by roughly 1.5 kcal/mol relative to the CCSD(T) barrier of approximately 3.3 kcal/mol. Using at most 2151 and (remarkably) as few as 430 LCCSD(T) energies, a permutationally invariant  $\Delta$-machine learned PES is developed that has an H-atom transfer barrier in  agreement with the LCCSD(T) one of 3.5 kcal/mol.  Tunneling splittings are re-calculated using this new PES using both fixed-node diffusion Monte Carlo calculations and a 1d tunneling path, and these are compared to earlier ones obtained using the MP2-based PES.

%Qualitative comparisons with a recent transfer-learning neural network approach to improve a very similar MP2-based PES lead us to conclude that the present approach is more efficient, i.e., requires far fewer high-level energies, than that approach.  We regard this conclusion as tentative, pending more investigations of the TL-NN approach, which we anticipate seeing.

%\clearpage

\bibliography{refs}

%\section*{Author Contributions}
%CQ and PLH performed the calculations. CQ, PLH, RC, and AN developed the computational software needed for the fitting basis sets and $\Delta$-machine learning approach. JMB designed and led the research. All authors contributed to write the manuscript.

%\section*{Competing financial interests}
%The authors declare no competing financial interests.

%\section*{Data Availability}
%Data supporting the findings of this study are available from the corresponding authors upon reasonable request.

%\section*{Code Availability}
%The potentials reported in the paper are available in compressed folders as supporting information.
%The code employed to generate the fitting basis sets will be made publicly available.

\end{document}